\documentclass[twocolumn,preprint2]{aastex631}

\usepackage{epsfig}
\usepackage{diagbox}
\usepackage{multirow}
\usepackage{float}
\usepackage[section]{placeins}

\begin{document}

\title{Evolution of the post merger remnants from the coalescence of oxygen-neon and carbon-oxygen white dwarf pairs}

\author[0000-0002-2452-551X]{Chengyuan Wu}
\affiliation{Yunnan Observatories, Chinese Academy of Sciences, Kunming 650216, China}
\email{wuchengyuan@ynao.ac.cn}

\author{Heran Xiong}
\affiliation{Research School of Astronomy and Astrophysics, The Australian National University, Canberra, ACT 2611, Australia}

\author{Jie Lin}
\affiliation{Physics Department and Tsinghua Center for Astrophysics,
Tsinghua University, Beijing, 100084, People’s Republic of China}

\author{Yunlang Guo}
\affiliation{Yunnan Observatories, Chinese Academy of Sciences, Kunming 650216, China}
\affiliation{University of Chinese Academy of Sciences, Beijing 100049, China}

\author{Xiaofeng Wang}
\affiliation{Physics Department and Tsinghua Center for Astrophysics,
Tsinghua University, Beijing, 100084, People’s Republic of China}

\author{Zhanwen Han}
\affiliation{Yunnan Observatories, Chinese Academy of Sciences, Kunming 650216, China}

\author{Bo Wang}
\affiliation{Yunnan Observatories, Chinese Academy of Sciences, Kunming 650216, China}

\begin{abstract}

Although multidimensional simulations have investigated the processes of double WD mergers, post-merger evolution only focused on the carbon-oxygen (CO) WD or helium (He) WD merger remnants. In this work, we investigate for the first time the evolution of the remnants stemmed from the merger of oxygen-neon (ONe) WDs with CO WDs. Our simulation results indicate that the merger remnants can evolve to hydrogen- and helium-deficient giants with maximum radius of about 300${R}_{\odot}$. Our models show evidence that merger remnants more massive than $1.95{M}_{\odot}$ can ignite Ne before significant mass-loss ensues, and they thus would become electron-capture supernovae (ECSNe). However, remnants with initial masses less than $1.90{M}_{\odot}$ will experience further core contraction and longer evolutionary time before reaching at the conditions for Ne-burning. Therefore their fates are more dependent on mass-loss rates due to stellar winds, and thus more uncertain. Relatively high mass-loss rates would cause such remnants to end their lives as ONe WDs. Our evolutionary models can naturally explain the observational properties of the double WD merger remnant $-$ IRAS 00500+6713 (J005311). As previously suggested in the literature, we propose and justify that J005311 may be the remnant from the coalescence of an ONe WD and an CO WD. We deduce that the final outcome of J005311 would be a massive ONe WD rather than a supernova explosion. Our investigations may be able to provide possible constraints on the wind mass-loss properties of the giants which have CO-dominant envelopes.

\end{abstract}


\section{Introduction} \label{sec:intro}

White dwarfs (WDs) are the most common evolutionary fate of single stars with initial masses lower than $9-11{M}_{\odot}$ (e.g. \citealt{1991ApJ...367L..19N}; \citealt{1999ApJ...515..381R}; \citealt{2004MNRAS.353...87E}; \citealt{2007A&A...476..893S}; \citealt{2007PhDT.......212P}; \citealt{2013ApJ...765L..43I}; \citealt{2017PASA...34...56D}; \citealt{2010A&A...512A..10S}). It is estimated that there are about $10^{10}$ WDs in our Galaxy out of which $2.5\times{10}^{8}$ reside in binaries consisting of two WDs (e.g. \citealt{2009JPhCS.172a2004N}; \citealt{2001A&A...365..491N}; \citealt{2009JPhCS.172a2022H}). About half of these binary WDs are close enough to begin mass transfer within Hubble time. 

The theoretical understanding of the evolution of WD merger remnants is based on using of information from three-dimensional (3D) post-merger configurations to build one-dimensional (1D) models, whose further evolution is computed in detail through their longer-duration phases. This procedure has been applied a number of times in the literature. For instance, \cite{2014MNRAS.445..660Z} investigated the merger of CO WD and He WD, and found that the merger remnant can evolve to R CrB stars. \cite{2022MNRAS.512.2972W} investigated the post-merger evolution of massive CO WD with He WD, and found that these systems contribute to a portion of ultra-massive CO WDs. \cite{2017ApJ...850..127B} investigated the merger of ONe WD with He WD, and found that these systems are related to electron capture supernovae (ECSNe). \cite{2016MNRAS.463.3461S} investigated double CO WD mergers (see also \citealt{2021ApJ...906...53S}), and found that the corresponding super-Chandrasakha-mass remnants can undergo iron-core collapse supernova explosions (Fe-CCSN). However, no post-merger evolutionary models has been built so far to study ONe WD + CO WD. For the evolutionary outcomes of such systems, although previous works suspected that they may related to ECSNe, SNe Iax or NSs (e.g. \citealt{1991ApJ...367L..19N}; \citealt{2006A&A...450..345K}), but detail investigates are desperately needed.

In this work, based on the results of three-dimensional (3D) simulations from \cite{2014MNRAS.438...14D}, we construct the 1D structures of ONe+CO WD merger remnants and investigate their evolution. We will show that the final outcomes of such remnants are more complicated than previous deductions, which could be massive ONe WD, ONeFe WD or NS depending on the mass of the merger remnant, the wind mass-loss process and the effects of the treatment of convection. Besides, our model can perfectly explain the main observational properties of object J005311 (e.g. \citealt{2019Natur.569..684G}; \citealt{2020A&A...644L...8O}; \citealt{2022arXiv220803946L}), which may provide a strong constrain on the wind mass-loss process of such merger remnants. The article is organized as follows. In Sect.\,2, we introduce our initial models of merger remnants. The evolution process of the remnants are shown in Sect.\,3. The discussion of model uncertainties are given in Sect.\,4. We provide our analyze and final fate of J005311 in Sect.\,5. The summary are given in Sect.\,6.

\section{Initial models} \label{sec:models}

\cite{2014MNRAS.438...14D} performed a large set of WD merger simulations, including 225 pairs of double WD mergers. The mass range of ONe WDs and CO WDs in their grid are from $1.10$ to $1.20{M}_{\odot}$ and from $0.65$ to $1.05{M}_{\odot}$ spaced at mass interval of $0.05{M}_{\odot}$, respectively. Almost all the mergers in their simulations are able to form stable remnants, except the double WD pairs with extremely low mass ratio (low mass He WDs merger with ONe WDs), which do not show dynamical mass-transfer during the merger. For the more massive merger remnant, \cite{2018ApJ...869..140K} simulated the process of $1.20{M}_{\odot}$ ONe WD merger with $1.10{M}_{\odot}$ CO WD, and found that the merger yields a failed detonation which may produce a very faint and rapidly fading transient. In the present work, we only consider the post-merger evolution of stable merger remnants (i.e. \citealt{2014MNRAS.438...14D}).

According to \cite{2014MNRAS.438...14D}, a double WD merger will form a remnant composed of a cool core, a hot envelope, a Keplerian disk and a tidal tail from the inside out. Based on the results of 3D simulations, they provided 1D remnant structure profiles, including key information of the merger remnants, such as their maximum temperature (${T}_{\rm max}$), the density at the location of the maximum temperature (${\rho}_{\rm max}$), their core mass (${M}_{\rm core}$), their envelope mass (${M}_{\rm env}$), and their disk mass (${M}_{\rm disk}$), among others. By considering the structures of merger remnants from \cite{2014MNRAS.438...14D}, \cite{2016MNRAS.463.3461S} developed the energy injecting prescription to map the initial configurations of the remnants and investigated the evolution of double CO WD mergers (see also \citealt{2021ApJ...906...53S}). 

In this work, we use a prescription similar to the one described in \cite{2016MNRAS.463.3461S} to construct a series of ONe WD+CO WD merger remnants with ONe WD mass in the range from 1.1 to $1.2{M}_{\odot}$ and total mass from 1.60 to $2.25{M}_{\odot}$ spaced at mass interval of $0.05{M}_{\odot}$ as our fiducial models by employing stellar evolutionary code {\tt\string MESA} (version 12778; e.g. \citealp{2011ApJS..192....3P, 2013ApJS..208....4P, 2015ApJS..220...15P, 2018ApJS..234...34P, 2019ApJS..243...10P}). In addition, we also constructed double CO WD merger remnants to compare with the observations (detailed information can be found in Sect.\,5). Since the mass of tidal tail only occupies a significantly small portion of total mass of the remnants, we ignore the influence of tidal tail and construct the merger remnants consisting only of a core, an envelope and a disk. As proved by \cite{2021ApJ...906...53S}, this approach has almost negligible influence on the initial profile of the remnants.

The detail prescriptions in constructing initial models are presented as follows. First, we create a series of pre-main sequence stars with masses equal to the total masses of the merger remnants, and evolve these models with nuclear reactions and elemental mixing turned off, until the central density reached ${10}^{3}\,{\rm g}\,{\rm cm}^{-3}$. Secondly, we resume the evolution, using the built-in capability of MESA to relax our model to a specified composition. In \cite{2014MNRAS.438...14D}, the composition of ONe WDs include 60\% of $^{\rm 16}{\rm O}$, 35\% of $^{\rm 20}{\rm Ne}$, 5\% of $^{\rm 24}{\rm Mg}$, and CO WDs include 40\% of $^{\rm 12}{\rm C}$ and 60\% of $^{\rm 16}{\rm O}$. Thus, the merger remnants in our models consist of an ONe core and an outer CO envelope. Once the composition relaxation is achieved, we evolve these models until their central densities reach ${10}^{5}\,{\rm g}\,{\rm cm}^{-3}$.

In order to construct the thermal structures of the merger remnants, we inject energy into different mass zones to alter the temperature and density until they reach target profiles, i.e. an isothermal core with a temperature of ${10}^{8}\,{\rm K}$, surrounded by a hot envelope and a disk. The target profiles are as follows: 

For $0\leq{M}_{\rm r}\leq{M}_{\rm {core}}$,
\begin{equation}
    T({M}_{\rm r})={T}_{\rm {core}}.
\end{equation}
For ${M}_{\rm {core}}\leq{M}_{\rm r}\leq{M}_{\rm {core}}+{M}_{\rm {env}}$,
\begin{equation}
    s({M}_{\rm r})={s}_{\rm {core}}+[{s}_{\rm {env}}-s({M}_{\rm {core}})]\frac{{M}_{\rm {r}}-{M}_{\rm {core}}}{{M}_{\rm {peak}}}.
\end{equation}
And for ${M}_{\rm {core}}+{M}_{\rm {env}}\le{M}_{\rm r}\leq{M}_{\rm {tot}}$,
\begin{equation}
    s({M}_{\rm r})={s}_{\rm {env}}.
\end{equation}
In the equations, ${M}_{\rm {core}}$, ${M}_{\rm {env}}$ and ${M}_{\rm {peak}}$ are, respectively, core masses, envelope masses and the mass coordinates of the maximum temperature of the merger remnants, which are related to the total masses of the remnants (${M}_{\rm {tot}}$) and the mass ratios of the double WDs before mergers ($q$), i.e.,
\begin{equation}
    {M}_{\rm {core}}={M}_{\rm {tot}}(0.7786-0.5114q),
\end{equation}
\begin{equation}
    {M}_{\rm {env}}={M}_{\rm {tot}}(0.2779-0.464q+0.716{q}^{2}),
\end{equation}
\begin{equation}
{M}_{\rm {peak}}={M}_{\rm {tot}}(0.863-0.3335q).
\end{equation}
In equation (1), ${T}_{\rm {core}}={10}^{8}\,{\rm K}$ is the typical temperature of the massive WD which used in our simulations. ${M}_{\rm peak}$ in equation (2) represents . ${s}_{\rm {core}}$ and ${s}_{\rm {env}}$ in equations (2) and (3) are, respectively, the specific entropies of the core and the envelope. We set ${s}_{\rm {core}}$ to be the specific entropy of the core when ${T}_{\rm {core}}={10}^{8}\,{\rm K}$, and ${s}_{\rm {env}}$ equals to the following equation:
\begin{equation}
    {\rm log}({s}_{\rm {env}}/{\rm {erg}}\,{\rm {g}^{-1}}\,{\rm {K}^{-1}})=8.7+0.3({M}_{\rm {tot}}/{M}_{\odot}-1.5).
\end{equation}
We use a very short time step (in seconds) to complete the energy injection process to guarantee the corresponding process cannot change the target profile (for more detail of the method, see \citealt{2016MNRAS.463.3461S} and \citealt{2021ApJ...906...53S}).

After the energy injection process, we obtained the initial structures of the merger remnants. Fig.\,1 shows the temperature-mass coordinate (${T}$-${M}_{\rm r}$) and temperature-density (${T}$-${\rho}$) profiles of different merger remnants after the energy injection processes. Our initial remnant models have isothermal cores at ${10}^{8}\,{\rm K}$. ${M}_{\rm env}$ and ${T}_{\rm max}$ of the merger remnants increase with increasing mass ratio of the progenitor double WDs and their total masses, which is consistent with 3D simulations.

\begin{figure*}
\begin{center}
\epsfig{file=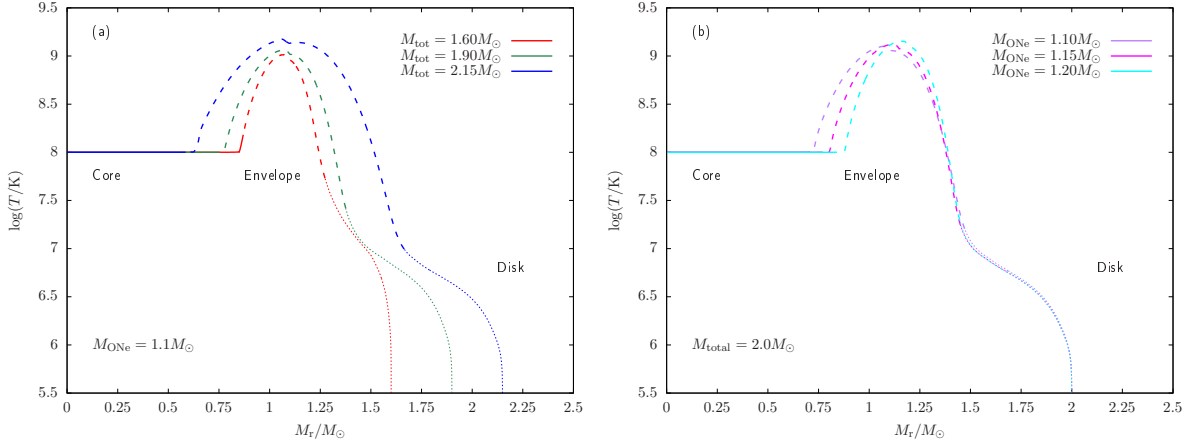,angle=0,width=16.2cm}
 \caption{Temperature-Mass coordinate profile of the merger remnants just after the energy injection processes. Panel (a): temperature profiles of three merger remnants of different total masses and the same ONe WD mass. Panel (b): temperature profiles of three merger remnants of equal ONe WD mass and different total masses. Different lines in panel (a) and (b) represent the corresponding structures of double WD mergers with various mass ratios.}
  \end{center}
  \label{fig: 1}
\end{figure*}

The comparison of some major parameters of our models with 3D simulations are presented in Fig.\,2. Although ${T}_{\rm max}$ and mass coordinate show slightly different with 3D simulations, our models conform to the variation trend of different merger remnants. Note, however, that the subsequent evolution of the remnants may be sensitive to the structures of the initial models. We analyse the influence of the initial conditions on the evolution of merger remnants in sect.\,4.1.

\begin{figure*}
\begin{center}
\epsfig{file=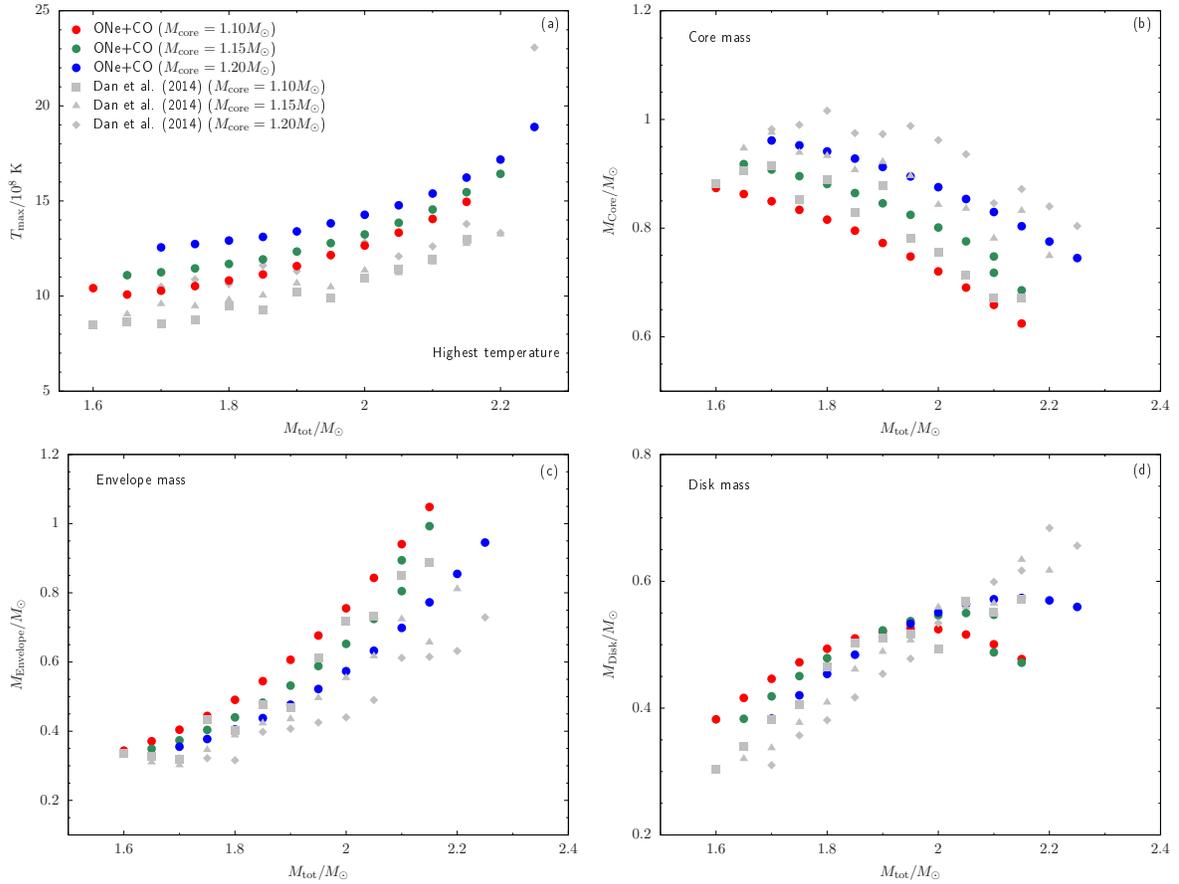,angle=0,width=16.2cm}
 \caption{Main properties of the merger remnants (${T}_{\rm max}$ in panel a; core mass in panel b; envelope mass in panel c; disk mass in panel d) with different initial ONe WD masses and CO WD masses. Red, sea-green and blue filled circles in each panel represent parameters of merger remnants with initial ONe WD masses equal to $1.10{M}_{\odot}$, $1.15{M}_{\odot}$ and $1.20{M}_{\odot}$ in our models, whereas grey diamonds in each panel represent the corresponding parameters resulted from \cite{2014MNRAS.438...14D}.}
  \end{center}
    \label{fig: 2}
\end{figure*}


\section{Evolution of the remnant of ONe WD+CO WD mergers} \label{sec:evolution}

\subsection{Input physics}

Since the outer layers of such merger remnants consist of carbon and oxygen, we use OPAL Type 2 opacity table to calculate the evolution of the remnants (e.g. \citealt{1996ApJ...464..943I}). When the remnants evolve to giant phase, the surface temperature may decrease to about ${\rm log}(T/{\rm K})=3.65$. Since the lower boundary of the OPAL tabulations is ${\rm log}(T/{\rm K})=3.75$, as in \cite{2016MNRAS.463.3461S}, we adopt the Kasen opacities to deal with the low temperature conditions, and blend the OPAL and Kasen opacities between ${\rm log}({T}/{\rm K})=4.1$ and $4.2$. This supplement is suitable for H- and He-deficient cool giant-like stars.

The nuclear reaction network used in the present work is ``${\tt {wd\_aic.net}}$'', which includes ``${\tt {co\_burn.net}}$'' in ``${\tt {MESA\,\,default}}$'', and we artificially added four isotopes (i.e. $^{\rm 20}{\rm O}$, $^{\rm 20}{\rm F}$, $^{\rm 24}{\rm Ne}$ and $^{\rm 20}{\rm Na}$) and the corresponding weak reactions in order to calculate the electron-capture process. The weak interaction rates are from \cite{2015MNRAS.453.1910S}. We apply ``${\tt {MESA\,\,default}}$'' screening factors to correct nuclear reaction rates for plasma interactions (e.g. \citealt{2007PhRvD..76b5028C}). The cooling rates from thermal neutrinos are derived from the fitting formulae of \cite{1996ApJS..102..411I}.

As the remnant expands toward giant phase, the superadiabatic gradient arising in radiation-dominated envelopes can force the adoption of prohibitively short time step. To circumvent this issue, we apply the treatment of convective energy transport ``MLT++'' (e.g. \citealt{2013ApJS..208....4P}), to reduce the superadiabaticity in the radiation-dominated convective regions. At soon as Ne burning is ignited under the CO envelope, the inwardly propagating flame will leave a thin burning shell with extremely high energy release rate. Previous works found that the treatment of convection can affect the evolution of the stars, for example, a small amount of convective boundary mixing can remove the physical conditions required for the off-center burning flame in super-AGB stars (e.g. \citealt{2013ApJ...772...37D}; \citealt{2014ApJ...797...83J}; \citealt{2015ApJ...807..184F}; \citealt{2016MNRAS.455.3848J}). In the present work, we consider overshooting process with parameters of ${\rm f}=0.014$ and ${\rm f}_{\rm 0}=0.004$ to calculate the inwardly propagating Ne flame  (e.g. \citealt{2013ApJ...772...37D}; \citealt{2020MNRAS.495.1445W}). We constrain the time resolution in our calculations by adopting ``${\tt{varcontral=3d-4}}$''. When the increasing central density of the remnant reached $\approx$ ${10}^{9.5}\,{\rm g}\,{\rm {cm}^{-3}}$, we constrain the time step by controlling the change of central density (${\tt {delta\_lgRho\_cntr\_limit=1d-3}}$) to guarantee the precision in simulating the electron-capture process. In order to have a better treatment on the propagation of off-center flame, we control the spatial resolution by changing ``${\tt {dlog\_burn\_c\_dlogP\_extra=0.1}}$'' and ``${\tt {dlog\_burn\_ne\_dlogP\_extra=0.1}}$'' to increase the mesh grid around the flame.

Some physical mechanisms have not been included in our fiducial models, such as wind mass-loss process, Urca process-cooling, stellar rotation, etc. We will discuss the effects of the corresponding uncertainties in sect.\,4.

\subsection{Post-merger evolution}

After constructing the initial models of the merger remnants, we restored the nuclear reactions and mixing process to let the remnants evolve forward in time. The detailed physical inputs of our fiducial models can be found in Table.\,2. The evolutionary features of two representative examples (a $1.1{M}_{\odot}$ of ONe WD merges with a $0.8{M}_{\odot}$ of CO WD, ${M}_{\rm {tot}}=1.90{M}_{\odot}$; a $1.1{M}_{\odot}$ of ONe WD merges with a $0.85{M}_{\odot}$ of CO WD, ${M}_{\rm {tot}}=1.95{M}_{\odot}$) are presented in Fig.\,3. Owing to the initially compact conﬁgurations, the remnants experiences an extremely short thermal adjustment stage (8hrs) prior to the ignition of carbon. Carbon burning starts below the convective envelope (the CO envelope is partial convective because of the energy injection process), proceeds through thermal pulses and forces the formation of an inner convective zone due to the high energy release. Nuclear energy is then transformed into work of expansion on the envelope, resulting in the increase of radius and opacity. As the temperature gradient increases, the convective zone develops gradually in the CO envelope. After the remnants approach the giant phase, carbon burning proceed steadily on the bottom of the shell, resulting in the mass increase of the ONe core. During the giant phase, the remnants can expand to about $300{R}_{\odot}$ in tens of years after merger. This giant stage takes the longest part of the evolution between the merger and the end of our calculations.

Carbon burning becomes stationary after some thermal pulses. The steady carbon burning leads to the mass increase of the degenerate core. The evolution of such giant-like object complies with core mass-luminosity relation (e.g. \citealt{1988MNRAS.235.1287J}). The major difference between $1.90{M}_{\odot}$ and $1.95{M}_{\odot}$ cases lays in the time when the Ne ignition occurs.

For the less massive remnants, carbon burning cannot triggers the Ne ignition immediately after the merger, and the remnants' cores need to experience longer evolutionary times to increase in mass. In this case, the ignition of Ne occurs after the electron capture process of $^{\rm 24}{\rm Mg}$ (i.e. 3015 years after merger) when the temperature and density of the Ne-burning region reaches ${\rm log}({T}/{\rm K})=9.11$ and ${\rm log}({\rho}/{\rm g}\,{\rm {cm}^{-3}})=7.23$, respectively. The mass coordinate of Ne ignition is at ${M}_{\rm r}=1.368{M}_{\odot}$, just below the carbon burning shell (${M}_{\rm r}=1.388{M}_{\odot}$, which corresponds to the mass of Ne core).

For the massive case, Ne ignition occurs $209$ years after the merger, at ${M}_{\rm r}=1.186{M}_{\odot}$ (relatively far away from the carbon burning shell, located at ${M}_{\rm r}=1.360{M}_{\odot}$), when the temperature and density of the Ne-burning region reaches ${\rm log}({T}/{\rm K})=9.23$ and ${\rm log}({\rho}/{\rm g}\,{\rm {cm}^{-3}})=7.37$, respectively. This means that the massive remnant does not experience a significantly long core increase phase, which leads to shorter evolutionary times of the remnants.

The major difference between these two cases can also be reviewed from the elemental abundance distributions (panel c and d of Fig.\,3). Owing to that the position of Ne ignition is very close to the CO burning shell in $1.90{M}_{\odot}$ case, the convective region caused by Ne burning penetrates into the CO envelope, resulting in the dredge-up of ONe-rich material out to the envelope. The abundance of Ne in CO envelope can reach about $30\%$. Meanwhile, note that the thickness of Si mantle is equals to the width of convective region caused by the Ne burning, therefore, the Si mantles in two cases are different because of the different positions of Ne ignitions.

\begin{figure*}
\begin{center}
\epsfig{file=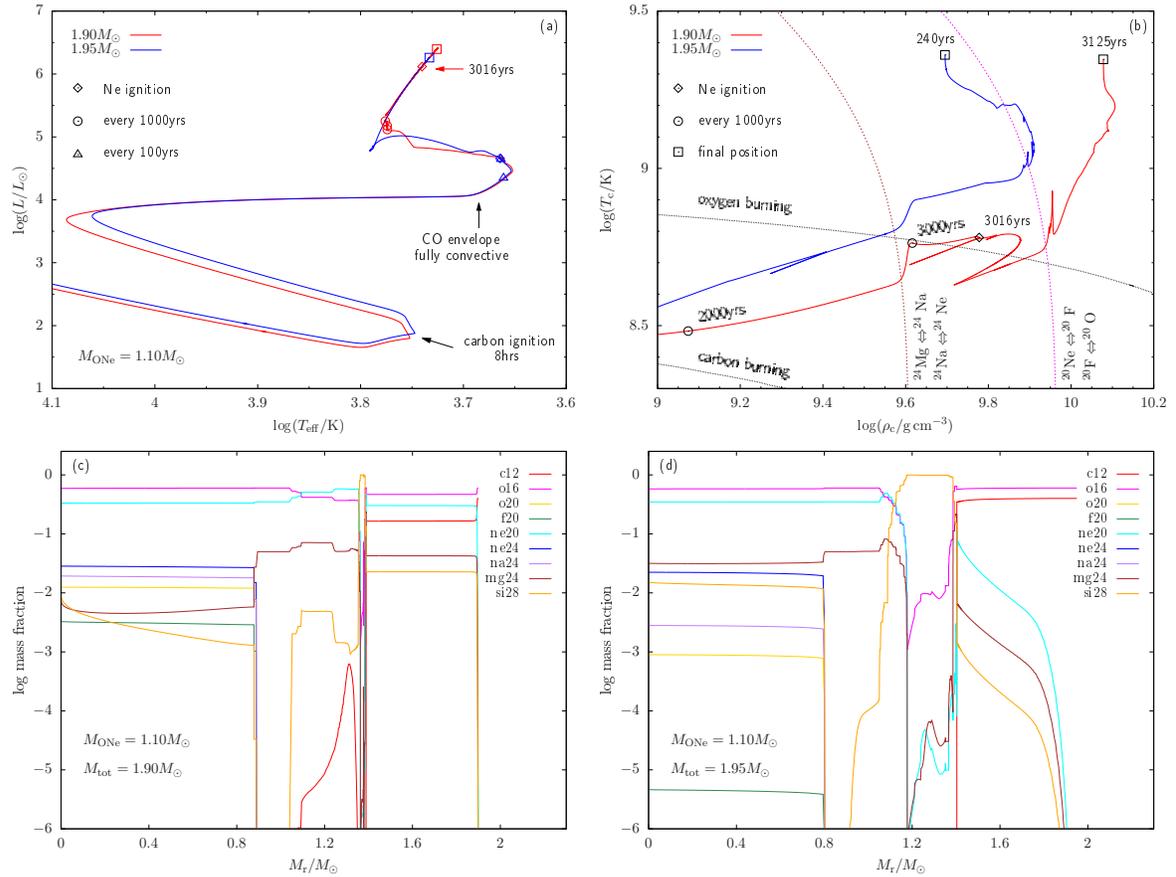,angle=0,width=16.2cm}
 \caption{An example of the evolution of two merger remnants (a $1.1{M}_{\odot}$ of ONe WD merges with a $0.8{M}_{\odot}$ of CO WD, ${M}_{\rm {tot}}=1.90{M}_{\odot}$; a $1.1{M}_{\odot}$ of ONe WD merges with a $0.85{M}_{\odot}$ of CO WD, ${M}_{\rm {tot}}=1.95{M}_{\odot}$). Panel (a): Hertzsprung–Russell diagram; panel (b): Central density-temperature evolution; panel (c) and (d): elemental abundance distributions of $1.90{M}_{\odot}$ and $1.95{M}_{\odot}$ merger remnants at the final stage of their evolution just prior to the explosions. In panels (a) and (b), diamonds represents the position of Ne ignition, whereas circles and triangles represent every 1000 years and 100 years after merger, respectively. Squares in panel (b) represents the position of final evolutionary stage, nearby which the corresponding evolutionary time are labeled. Black dotted lines in panel (b) represents the carbon and oxygen burning lines, whereas brown and magenta dotted lines represent starting lines of electron capture reactions of $^{\rm 24}{\rm Mg}$ and $^{\rm 20}{\rm Ne}$. Note that diamond on blue line of panel (b) is out of range.}
  \end{center}
    \label{fig: 3}
\end{figure*}

In addition to the above-mentioned evolutionary times, the final densities of the two cases considered also show differences. To better analyse these differences, we present the Hertzsprung–Russell diagram (HRD) and central density-temperature profiles for a series of remnants in Fig.\,4. Although the evolutionary processes follow similar trends on HRD, they show differences in ${T}_{\rm c}$-${\rho}_{\rm c}$ profile. Due to the different ignition times of Ne mentioned above, the corresponding two classes of remnants (${M}\leq1.90{M}_{\odot}$ and ${M}\geq1.95{M}_{\odot}$) can evolve to different central densities when explosion occurs. Previous multidimensional dynamical simulations of supernova explosion suggested that for the pure ONe core, Ne or O explosion occur at higher central density (${\rm log}({\rho}_{\rm c}/{\rm g}\,{\rm cm}^{-3})>9.97$) will result in the formation of NS, whereas for the cases of which Ne or O explosion occur at lower central density (${\rm log}({\rho}_{\rm c}/{\rm g}\,{\rm cm}^{-3})<9.97$), a ONeFe core would leave behind after the explosion (e.g. \citealp{2016A&A...593A..72J, 2019A&A...622A..74J}). Our results indicate that the final outcomes of double WD mergers strongly depend on the initial remnant masses.

\begin{figure*}
\begin{center}
\epsfig{file=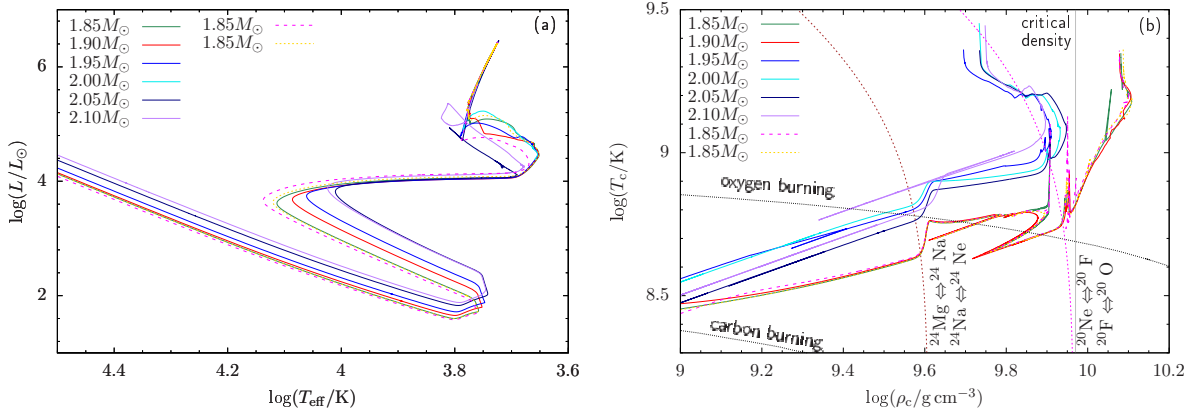,angle=0,width=16.2cm}
 \caption{Evolutionary tracks of different remnants (${M}_{\rm ONe}^{\rm i}=1.10{M}_{\odot}$) in HRD and in central temperature-central density profile. Black dotted lines in panel (b) represent the carbon and oxygen burning conditions, whereas brown and magenta dotted lines represent starting lines of electron capture reactions of $^{\rm 24}{\rm Mg}$ and $^{\rm 20}{\rm Ne}$. Grey vertical line in panel (b) represents the critical density below which the explosion of the core will lead to the formation of ONeFe WD, whereas above which will lead to the formation of NS.}
  \end{center}
    \label{fig: 4}
\end{figure*}

\section{Model uncertainties}

Previous studies indicate that some pivotal physical mechanisms such as wind mass-loss process can significantly affect the evolution of double WD merger remnants (e.g. \citealt{2019ApJ...885...27S}; \citealt{2021ApJ...906...53S}; \citealt{2022MNRAS.512.2972W}). We discuss the influence of these mechanisms in the current section.

\subsection{Initial structure}

The initial structures of the merger remnants of our models result from the fitting formulae in \cite{2014MNRAS.438...14D}. The comparison of some major parameters of our models with those from 3D simulations are presented in Fig.\,5. Although ${T}_{\rm max}$ and mass coordinate show slightly different with 3D simulations, our models conform to the variation trend of different merger remnants. However, the fitting formulae for different properties of merger remnants in \cite{2014MNRAS.438...14D} can only explain 75\% to 90\% of the variance. In order to investigate the effects of initial model uncertainties, we constructed models with different ${T}_{\rm max}$ by changing the entropy of the envelope (${s}_{\rm env}$ in Eq.\,4). Basically, a higher envelope entropy can lead to a lower ${T}_{\rm max}$ because of the decreased in density of envelope. On the contrary, a lower envelope entropy can result in a higher ${T}_{\rm max}$. Detail physical inputs are given in table.\,3.

The evolution of $1.90{M}_{\odot}$ and $1.95{M}_{\odot}$ merger remnants with different initial ${T}_{\rm max}$ are shown in Fig.\,5. Decreasing the envelope entropy (increasing ${T}_{\rm {max}}$) in our models shifts downwards the lower mass threshold for the formation of an ONeFe WD. For the less massive remnant (${M}=1.90{M}_{\odot}$), to increases the envelope entropy (a lower ${T}_{\rm max}$) has no influence in the final outcome, whereas increasing in ${T}_{\rm max}$ will lead to the similar evolutionary track as massive remnant. On the other hand, the outcome of our $1.95{M}_{\odot}$ model is not affected by an increase in ${T}_{\rm {max}}$, but delays Ne-ignition if ${T}_{\rm max}$ is decreased. The dependence of the outcomes with ${T}_{\rm max}$ reflects the influence of the initial remnant mass on the final fate of the considered models.

\begin{figure*}
\begin{center}
\epsfig{file=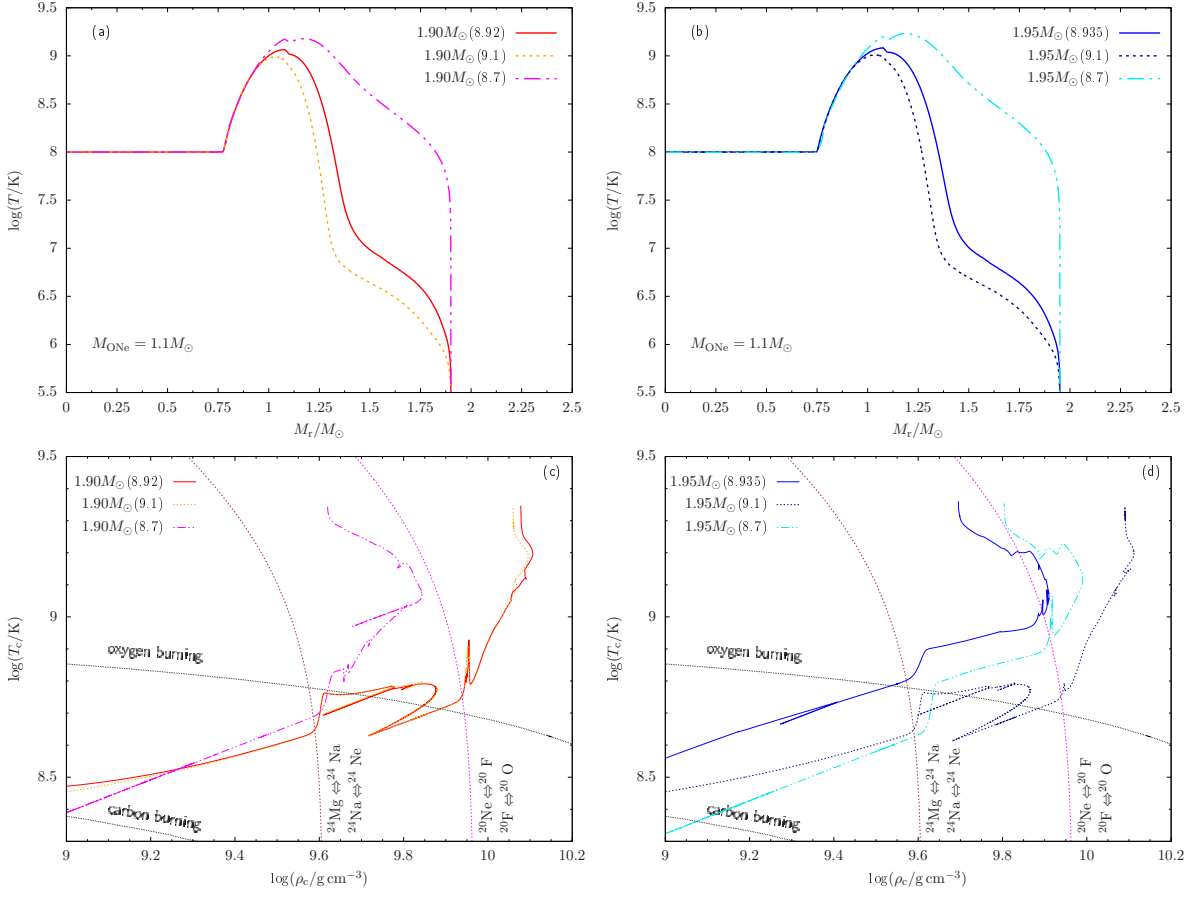,angle=0,width=16.2cm}
 \caption{Initial profiles and evolutionary tracks of $1.90{M}_{\odot}$ and $1.95{M}_{\odot}$ merger remnants with different envelope entropy. Panel (a) and (b): initial $T$-${M}_{\rm r}$ profiles. Red (blue) solid line represents our fiducial model, whereas magenta (cyan) dashed-dotted-dotted line and golden (navy) dotted line represent remnants with lower (${\rm ln}({s}_{\rm env}/{\rm erg}\,{\rm {g}^{-1}}\,{\rm {K}^{-1}})=8.7$) and higher (${\rm ln}({s}_{\rm env}/{\rm erg}\,{\rm {g}^{-1}}\,{\rm {K}^{-1}})=9.1$) envelope entropy, respectively. Panel (c) and (d): evolutionary tracks of merger remnants in ${\rm log}{T}_{\rm c}-{\rm log}{\rho}_{\rm c}$ diagram.}
  \end{center}
    \label{fig: 5}
\end{figure*}

\subsection{Mass-loss rate}

The wind mass-loss prescription has not been considered in our fiducial models. During their evolution, the merger remnants can evolve to giant phase and produce dusty winds, which may have influence on their evolution. However, the physics of the winds of such H-/He-deficient giants is not well understood, and mass-loss rates cannot be constrained from the observation so far. In the present work, we simply explore the effect of mass-loss by considering Reimers' (e.g. \citealt{1975MSRSL...8..369R}) and Blocker's (e.g. \citealt{1995A&A...297..727B}) wind mass-loss prescription with ``${\tt {MESA\,\,default}}$'' values of wind coefficients (${\eta}_{\rm R}=0.5$; ${\eta}_{\rm B}=0.05$). Detail physical inputs are given in Table.\,4.

Fig.\,6 presents the evolutionary tracks of $1.90{M}_{\odot}$ ($1.10$+$0.80{M}_{\odot}$) and $1.95{M}_{\odot}$ ($1.10$+$0.85{M}_{\odot}$) ONe+CO WD merger remnants under Reimers' and Blocker's wind mass-loss prescriptions. During the giant phase, the wind mass-loss rate with Reimers' wind prescription is significantly less than that under Blocker's wind prescription, which only exceeds the order of magnitude of ${10}^{-4}\,{M}_{\odot}\,{\rm yr}^{-1}$ at the end of the evolution. This means that Reimers' wind cannot significantly change the total masses of the remnants, and the final outcome of the remnants are same as those without wind mass-loss. However, under the Blocker's wind prescription, the remnants can maintain high mass-loss rate during the whole giant phase, which effectively decrease the total mass of the remnants. For the less massive remnants (${M}\leq1.90{M}_{\odot}$), all the CO envelope can be lost during the evolution, resulting in the formation of ONe WDs. Although the strong wind mass-loss process can obviously affect the massive remnants (${M}\geq1.95{M}_{\odot}$), because of the short evolutionary time, the final mass of the initially massive models can still be high enough to trigger the electron capture reactions that lead to the formation of ONeFe WDs. Our results indicate that wind prescriptions play a critical role in investigating the evolution of double WD merger remnants. Whether NSs can be formed though ONe+CO WD binary systems may strongly depend on the wind efficiency after the merger.

\begin{figure*}
\begin{center}
\epsfig{file=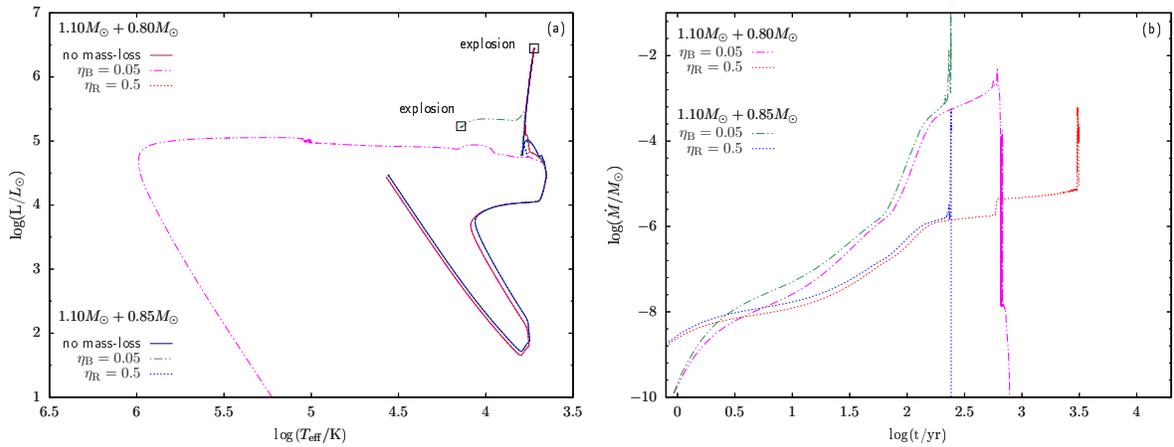,angle=0,width=16.2cm}
 \caption{Evolutionary of $1.90{M}_{\odot}$ and $1.95{M}_{\odot}$ ONe+CO WD merger remnants under different wind mass-loss prescriptions. Panel (a): HRD. The representative of each line are labeled. Squares on each line represent the positions on HRD of the corresponding models when supernova explosion occur.} Panel (b): evolution of mass-loss rate.
  \end{center}
    \label{fig: 6}
\end{figure*}

\subsection{Initial helium abundance}

According to the stellar evolutionary results, there exists a thin He shell outside the CO core for the low mass CO WDs. The 0.5 to $0.6{M}_{\odot}$ CO WDs in \cite{2014MNRAS.438...14D} models have $0.1{M}_{\odot}$ of He envelope. During the merger process, He is fully mixed with CO material and form the hot envelope surrounding the post-merger ONe WD. The exist of He in the merger remnant may influence the evolution of the remnants. To investigate the effect of the existence of He in our initial models, we build an alternative $1.60{M}_{\odot}$ ONe WD+CO WD merger remnant, by changing the composition described in Sect.\,2 to 40\% of $^{\rm 12}{\rm C}$, 40\% of $^{\rm 16}{\rm O}$ and 20\% of $^{\rm 4}{\rm He}$. Then we let this model evolve in time. Detailed physical inputs are given in Table.\,5.

In Fig.\,7, we present the evolutionary tracks of $1.60{M}_{\odot}$ ONe+CO WD (original composition) and the ONe+HeCO WD (alternative composition) merger remnants on the HRD. Due to the non-negligible amount of He in the envelope, the temperature required for the He-rich envelope burning is expected to be lower than that without helium. Thus, the convective envelope develops very fast in the He-rich envelope, it becomes fully convective in only 10 hours after merger, resulting in the difference during the luminosity increase phase. Furthermore, owing to the lower nuclear reaction rate in the He-rich burning shell comparing with the pure CO counterpart, the ONe+HeCO WD merger remnants need longer evolutionary time to increase the core mass until explosive Ne/O ignition occurs. Although the existence of He in the envelope influences the evolutionary time of the remnants, the final outcome are the same.

\begin{figure*}
\begin{center}
\epsfig{file=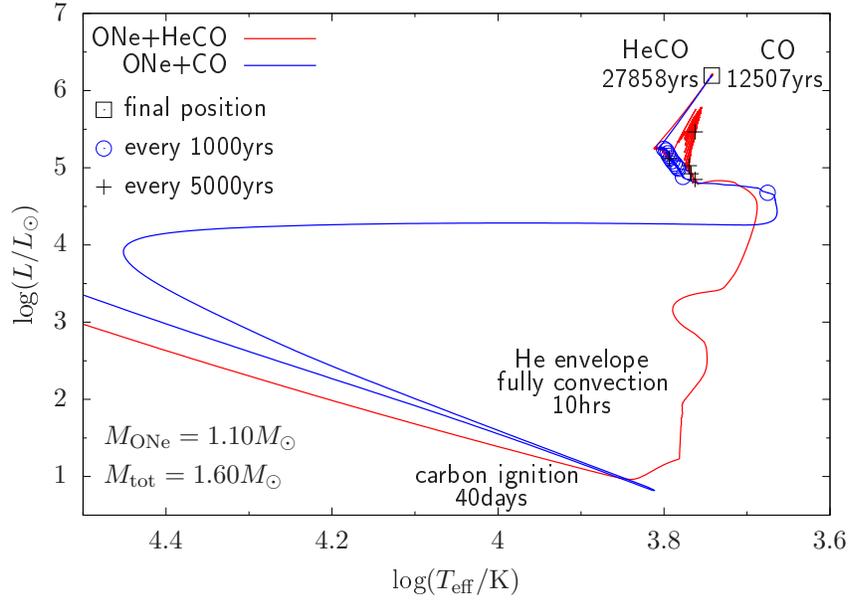,angle=0,width=12.2cm}
 \caption{Evolutionary tracks of $1.60{M}_{\odot}$ ONe+CO WD and ONe+HeCO WD  merger remnants on HRD. Black crosses on red curve represent every 5000 years during evolution, whereas blue circles on blue curve represent every 1000 years during evolution. Black square represents final position on HRD for each evolution track.}
  \end{center}
    \label{fig: 7}
\end{figure*}

\subsection{Urca-process cooling}

Urca-process-cooling may effect the central temperature during the evolution. We construct $1.90{M}_{\odot}$ and $1.95{M}_{\odot}$ merger remnants with 5\% of $^{\rm 23}{\rm Na}$ and 1\% of $^{\rm 25}{\rm Mg}$ in the initial ONe WD (e.g. \citealt{2017MNRAS.472.3390S}), and simulate their evolution to investigate the effect of Urca-process-cooling. Detail physical inputs are given in table.\,6.

From Fig.\,8, we can see that by considering Urca-process-cooling, the central temperature decreased during the stage of electron capture reaction of $^{\rm 25}{\rm Mg}$ and $^{\rm 23}{\rm Na}$, which is similar to what happened in mass-accretion WD systems (e.g. \citealt{2017MNRAS.472.3390S}). The remnants evolve faster when Urca-process-cooling is considered. This is because due to the electron capture of $^{\rm 25}{\rm Mg}$ and $^{\rm 23}{\rm Na}$, the central temperature of remnant decreased, resulting in the shrinking of the ONe core which promote the occurrence of electron capture reaction of $^{\rm 24}{\rm Mg}$ and $^{\rm 20}{\rm Ne}$. The examples show that the final fates of merger remnants do not changed when considering Urca-process cooling, which means that the corresponding effect is a minor factor in such merger remnants.

\begin{figure*}
\begin{center}
\epsfig{file=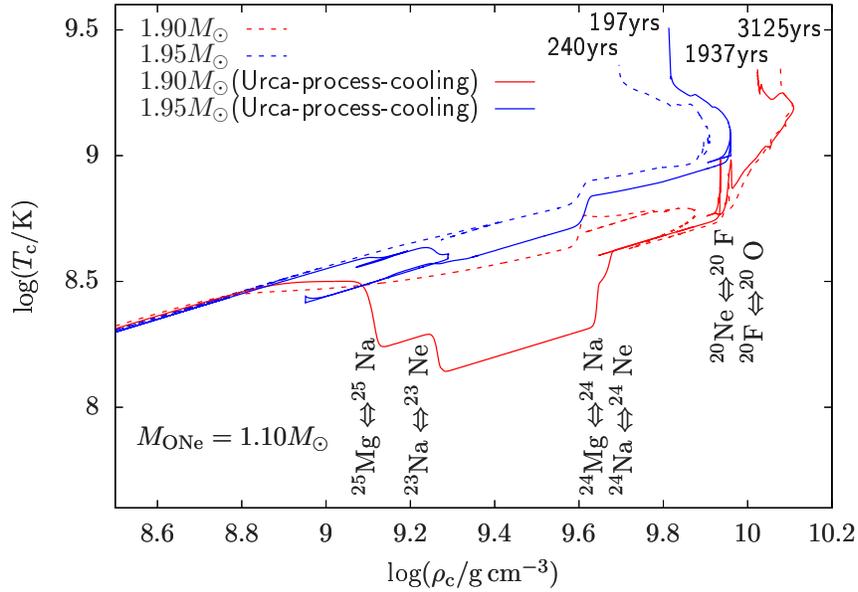,angle=0,width=12.2cm}
 \caption{Evolutionary tracks of $1.90{M}_{\odot}$ and $1.95{M}_{\odot}$ merger remnant in ${\rm log}{T}_{\rm c}-{\rm log}{\rho}_{\rm c}$ diagram. Solid and dashed lines represent the corresponding tracks when considering and without considering Urca-process-cooling during evolution, respectively.}
  \end{center}
    \label{fig: 8}
\end{figure*}

\subsection{Rotation}

During the merger process, a portion of orbital angular momentum could be left in the remnant. Rotation can decrease the density in the burning zone and enhance the mixing process, which may have influence on the evolution of the remnants.

In order to investigate the rotational effects, we inject torque into different mass zones in $1.90{M}_{\odot}$ and $1.95{M}_{\odot}$ initial models. 3D simulations predicted that the merger remnants is consist of a corotating core with ${\Omega}_{\rm c}\approx0.23-0.52\,{\rm rad}/{\rm s}$ (depended on total mass and mass ratio), and an differential rotating envelope of which ${\Omega}_{\rm e}$ gradually decreases from the inside out. In the present work, we consider two different rotational velocity. The fast rotational models have uniform angular velocity of ONe core (${\Omega}_{\rm c}=0.5\,{\rm rad}/{\rm s}$) and different rotational velocity in envelope (${\Omega}_{\rm e}=0.5\,{\Omega}_{\rm crit}$), whereas the slow rotational models have the same angular velocity distributions but with ${\Omega}_{\rm c}=0.25\,{\rm rad}/{\rm s}$, and ${\Omega}_{\rm e}=0.25\,{\Omega}_{\rm crit}$, respectively. Note that magnetic fields are quite effective in transporting angular momentum in double WD merger remnants, therefore, we adopt the angular momentum transport prescriptions described in \cite{2002A&A...381..923S} (ST) and \cite{2019MNRAS.485.3661F} (FPJ) to simulate the evolution of the remnants, respectively. Detail physical inputs are given in Table.\,7.

From Fig.\,9, we can see that the remnants computed with the FPJ prescription can transport angular momentum from the core to the envelope more rapidly. Thus, in this case, the envelope reaches the same angular velocity as the core has within hours after merger. As the remnants expand towards the giant phase, the angular velocity of the envelope obviously decreases. The opacity of the outer shell changes due to the variation of temperature and rotational mixing process, which leads to differences in the phase in which luminosity increase. Although the final fate of the rotational models are same as those of the fiducial models, the difference of core rotational velocity caused by the different prescriptions of angular momentum transport implies that the efficiency of angular momentum transport is one of the most important mechanism in influencing the rotational velocity of massive WDs.

\begin{figure*}
\begin{center}
\epsfig{file=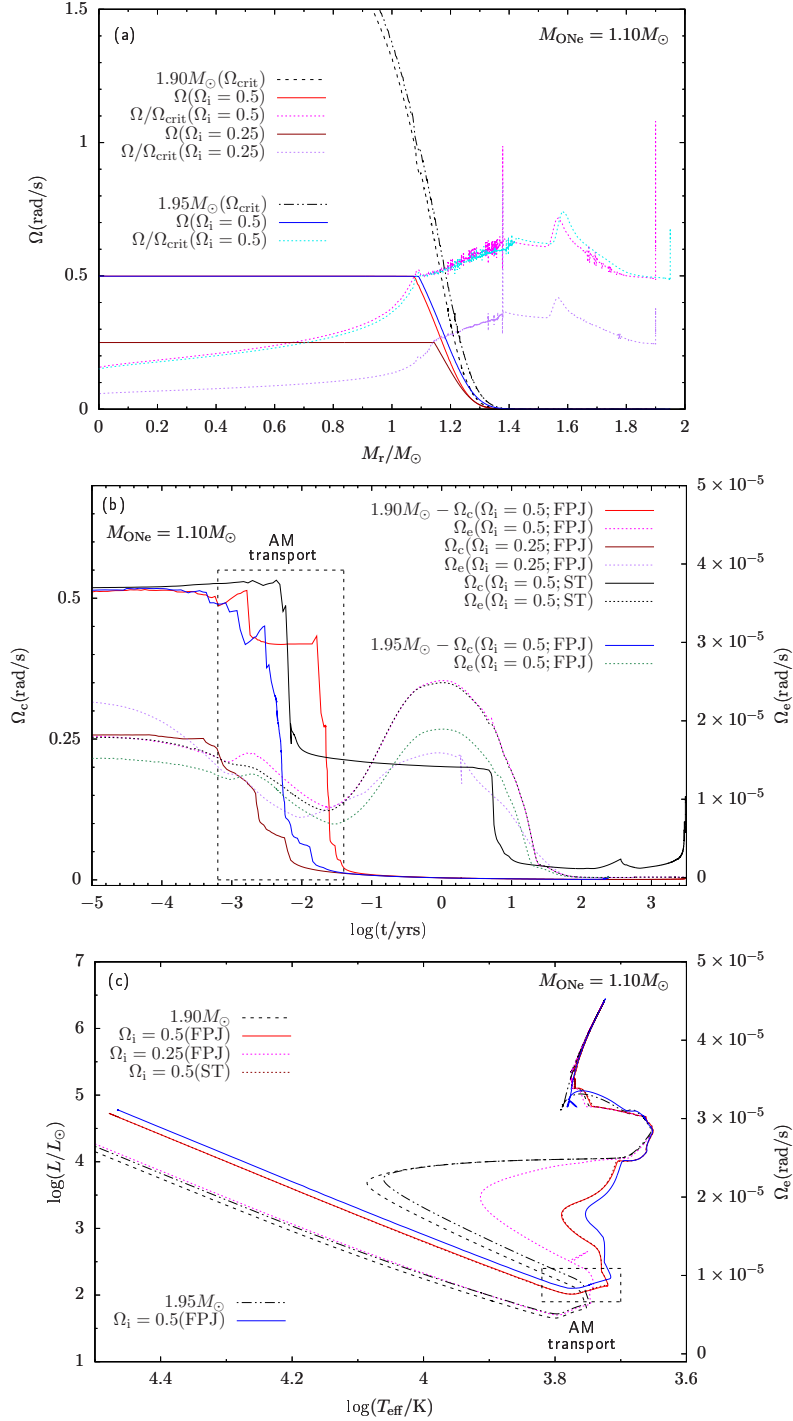,angle=0,width=11.2cm}
 \caption{Evolutionary tracks when consider rotational effect. Panel (a): initial rotation profile of $1.90{M}_{\odot}$ and $1.95{M}_{\odot}$ merger remnants. Solid, dashed/dashed-dotted-dotted and dotted lines represent angular velocity, critical angular velocity and the ratio of angular velocity and critical angular velocity of  $1.90$ and $1.95{M}_{\odot}$ merger remnants with different initial rotational velocities, respectively. Panel (b): evolution of core angular velocity (solid lines) and envelope angular velocity (dotted lines) of four models under different angular momentum transport prescriptions. Region inside the rectangle represents the most visible portion of the angular momentum transfer process. Panel (c): evolutionary tracks of four models comparing with non-rotational models on HRD. Similar to panel (b), region inside the rectangle represents the most visible portion of the angular momentum transfer process.}
  \end{center}
    \label{fig: 9}
\end{figure*}

\section{J005311}

Owing to the small quantity of massive double WD systems (e.g. \citealt{2020MNRAS.494.3422L}) and the extremely short evolutionary time of the corresponding merger remnants (e.g. \citealt{2021ApJ...906...53S}) , it is very difficult to observe such remnants. Fortunately, \cite{2019Natur.569..684G} discovered a hot star in the constellation Cassiopeia (RA = 00h 53 min 11.21s and dec. = +${67}^{\circ}$ 30' 2.1'', hereafter J005311), located at the center of a circular mid-infrared nebula by using data from the Wide-field Infrared Survey Explorer (WISE). The center star has V band magnitude of 15.5 mag and, according to GAIA DR2, is located at a distance of 3.06 Kpc. The nebula appears as a circular shell and a diffuse halo with linear radii of about 1.1pc and 1.6pc, respectively. The optical follow-up spectroscopy of J005311 obtained from the Russian 6-m telescope revealed an emission-line-dominated spectrum, similar to oxygen-rich Wolf-Rayet (WO type) star. This emission-line-dominated spectrum was later confirmed by \cite{2020RNAAS...4..167G} who used the Large Binocular Telescope (LBT) to obtained the spectra of J005311, and found significant variability in the profile of strong O\,VI emission feature which was predicted to be caused by rapidly shifting subpeaks generated by clumpiness in the stellar wind of J005311. By employing  the Potsdam Wolf-Rayet code, \cite{2019Natur.569..684G} analysed the optical spectrum of J005311, and found that the composition of the material at the base of wind was dominated by oxygen and carbon with mass fractions of 0.8$\pm$0.1 and 0.2$\pm$0.1, respectively. Based on the width and strength of the O\,VI emission line, they estimated that the wind mass-loss rate of the object was $\dot{M}=(3.5\pm0.6)\times{10}^{-6}\,{M}_{\odot}\,{\rm {yr}^{-1}}$ with a terminal wind velocity of ${v}_{\infty}=16000\pm1000{\rm {km}}\,{\rm {s}^{-1}}$. On the basis of these observations and the numerical results from \cite{2016MNRAS.463.3461S}, \cite{2019Natur.569..684G} argued that J005311 was the outcome of a super-Chandrasekhar-mass double CO WD mergers. Furthermore, they suggested that it would end its life as a Fe-CCSN.

\cite{2020A&A...644L...8O} reported on the first deep X-ray observation of J005311 and the first spectroscopic investigation of its nebula. They found that the central star is more luminous in X-ray than single massive OB and WR stars. Combining with optical data, they claimed that J005311 is a super-Chandrasekhar mass ONe WD+CO WD merger remnant. Similar to \cite{2019Natur.569..684G}, they suggested that J005311 would likely terminate its evolution as type I supernova, in which the final core collapse to a neutron star might be induced by electron captures process.

\cite{2021ApJ...918L..33R} later suggested that J005311 and its nebula are the residual core and ejecta of the 1181 AD explosion. They argued that this source is the result of an ONe+CO WD merger, and it is the only second type Iax supernova living in our Galaxy, which provided strong observational support for the double-degenerate merger scenario for type Iax supernovae. \cite{2022arXiv220803946L} modelled the stellar remnant of J005311, and derived a magnetic field with upper limit of 2.5MG, and the luminosity indicated a remnant mass of $1.2\pm0.2{M}_{\odot}$ with ejecta mass of $0.15\pm0.05{M}_{\odot}$. Combining with the low mass ratio of Ne and O (Ne/O$\,<0.15$), they suggested that the object is originated from the merger of double CO WDs or CO+He WDs, argued against the ONe WD merger origin of this object.

The origin of J005311 still remains uncertain. To understand the nature of J005311, we recalculated ONe WD+CO WD merger models by considering wind mass-loss and rotation process, since on one hand the observation supported that the object has extremely strong mass-loss process, and on the other hand, double WD merger remnants could keep a portion of orbital angular momentum. As in the prescription described in Sect.\,4.5, we inject torque into $1.6$ and $1.8{M}_{\odot}$ merger remnants to change their initial rotational profiles until their core (${\Omega}_{\rm c}$) and envelope (${\Omega}_{\rm e}$) rotational velocity reached ${\Omega}_{\rm c}=0.5\,{\rm rad}/{\rm s}$ and ${\Omega}_{\rm e}=0.5\,{\Omega}_{\rm crit}$ (${\Omega}_{\rm crit}$ is the Keplerian velocity for each mass zone). The corresponding rotational profile can approximately describe the properties of the merger remnants with similar total mass (e.g. \citealt{2009A&A...500.1193L}). During the post-merger evolution, we simply consider Blocker's wind mass-loss prescription (e.g. \citealt{1995A&A...297..727B}) with typical scaling factor of ${\eta}_{\rm B}=0.05$ to produce strong mass-loss process. In comparison, we also calculated CO WD+CO WD merger models under the same physical inputs.

Some key stellar parameters of J005311 and of our models are listed in Table\,1 and Table\,8, respectively. Fig.\,10 shows that the evolutionary tracks on the HRD of all these four models pass through a position close to J005311. However, the double CO WD merger model has very low mass-loss rates, in the range from ${10}^{-9}$ to a few of ${10}^{-8}\,{M}_{\odot}\,{\rm {yr}^{-1}}$, when the corresponding remnant has a temperature similar to that of J005311. The reason is that such merger remnant experiences off-center carbon burning through flashes, which prevents stationary shell burning near the edge of the degenerate core, resulting in a lower luminosity remnant. For the ONe WD+CO WD model, because of the initially low mass and the short evolutionary time of the remnant due to the strong wind mass-loss process, off-center neon burning cannot be triggered after merger. Thus, the remnant still maintain shell carbon burning, and has relatively high mass-loss rates comparable to the values inferred from observations.

\begin{table*}[htbp]
\centering
\begin{tabular}{|l|r|r|r|r|r|}
\hline
& Observation & \multicolumn{2}{c|}{ONe+CO WD} & \multicolumn{2}{c|}{CO+CO WD}\\\hline
Parameter & J005311 & 1.08+0.52 & 1.10+0.70 & 0.85+0.75 & 0.95+0.85 \\\hline
T (K) & $211000_{-23000}^{+40000}$ & $18800-25100$ & $18800-25100$ & $18800-25100$ & $18800-25100$ \\
L (${L}_{\odot})$ & $39810_{-10970}^{+20144}$ & $58606-59503$ & $76357-77520$ & $11000-14190$ & $16200-22527$ \\
$\dot{M}$ (${M}_{\odot}\,{\rm {yr}}^{-1}$) & $3.5(\pm0.6)\times{10}^{-6}$ & $3.2-6.06\times{10}^{-6}$ & $8.627-1.47\times{10}^{-6}$ & $3.2-7.65\times{10}^{-9}$ & $1.23-3.0\times{10}^{-8}$ \\
M (${M}_{\odot}$) & $1.2\pm0.2$ & $1.186$ & $1.252$ & $1.313$ & $1.463$\\
R (${R}_{\odot}$) & $0.15\pm0.04$ & $0.128-0.23$ & $0.148-0.26$ & $0.06-0.108$ & $0.08-0.122$ \\
age (yr) & unknown & $1586-1644$ & $1058-1082$ & $11371-18072$ & $12339-15863$ \\
final fate & unknown & ONe WD & ONe WD & ONe WD & Fe-CCSN\\
\hline
\end{tabular}
\caption{\label{tab:J005311} Major parameters (temperature, luminosity, mass-loss rate, stellar mass, stellar radius, stellar age and final fate) of J005311 from the observations (second column) and from our theoretical models when the remnants evolve to the same temperature as J005311 (columns 3-6).}
\end{table*}

Besides, results from double CO WD merger and ONe+CO mergers allow to infer different model remnant masses when the associated tracks reach the position of J005311 in the HRD. After experiencing extremely strong stellar wind, the masses of ONe WD+CO WD merger remnants decrease and reach values in the range from $1.186$ to $1.253{M}_{\odot}$. These sub-Chandrasekhar remnant mass values are consistent with those derived from observation, i.e. $1.2\pm0.2{M}_{\odot}$ (e.g. \citealt{2022arXiv220803946L}). For the double CO WD merger counterparts, the initially massive remnant (${M}^{\rm i}=1.80{M}_{\odot}$) is still super-Chandrasekhar mass ($1.473{M}_{\odot}$), which predict a different evolutionary outcomes for J005311, i.e. it will undergo supernova explosion through Fe-CCSN. The final fate of the less massive double CO merger remnant (${M}^{\rm i}=1.60{M}_{\odot}$) is the same as that of the ONe+CO WD merger of the same initial mass. However, the radius of the former is too small to explain the observed properties of J005311. Note that for the super-Chandrasekhar mass remnants, the deflagration of oxygen may produce ONeFe WD with mass close to $1.2{M}_{\odot}$ (e.g. \citealt{2016A&A...593A..72J}), which may also show similar characteristics as J005311. However, such post-explosion remnant should have extremely low carbon abundance (e.g. \citealt{2019A&A...622A..74J}), which may not be consistent with J005311. Hence, we suggest that J005311 derives from an ONe+CO WD merger and, in the future, it is expected to evolve to an ONe WD with a mass close to $1.20{M}_{\odot}$ (see also \citealt{2019ApJ...887...39K}).

\cite{2022arXiv220803946L} argued that J005311 originates from a double CO WD merger rather than from a ONe WD+CO WD merger, since the ejecta of J005311 is deficient in neon. Based on our simulations, the convective regions in the envelope of neither the ONe WD+CO WD remnant nor the double CO WD remnants are able to penetrate the carbon burning zone. This leads to expect neon deficient ejecta at the corresponding evolutionary stage of J005311 (see panel c and d of Fig.\,10). Note that owing to the occurrence of mixing process during the double WD merger, the base of CO envelope in the ONe+CO merger remnant could show neon enhancement. However, according to \cite{2014MNRAS.438...14D}, the abundance of neon at the base of the CO envelope cannot be higher than $20\%$, which may still be consistent with a low mass ratio of Ne and O as the remnant evolves to pre-WD.

Meanwhile, the nebula of J005311 is $1.1$ to $1.6$ pc far away from the central star. \cite{2021ApJ...918L..33R} argued that the nebula is stemmed from the supernova explosion of 1181 AD. If a NS is formed after the supernova explosion, it is difficult to explain why the central remnant remain constant luminosity within 100 years (e.g. \citealt{2020A&A...644L...8O}). Considering that the wind velocity of pre-WD is in the order of magnitude of $1000{\rm km}/{\rm s}$, it will spend $1120$ to $1600$ years to form the size of nebula of J005311, which is consist with the evolutionary time of ONe+CO WD merger remnant. Therefore, we suggest that J005311 may resulted from ONe WD+CO WD merger, and its nebula may formed though the strong wind-loss process during its evolution. In addition, we predict that this object will form a massive WD rather than undergo supernova explosion thousand of years later.

\begin{figure*}
\begin{center}
\epsfig{file=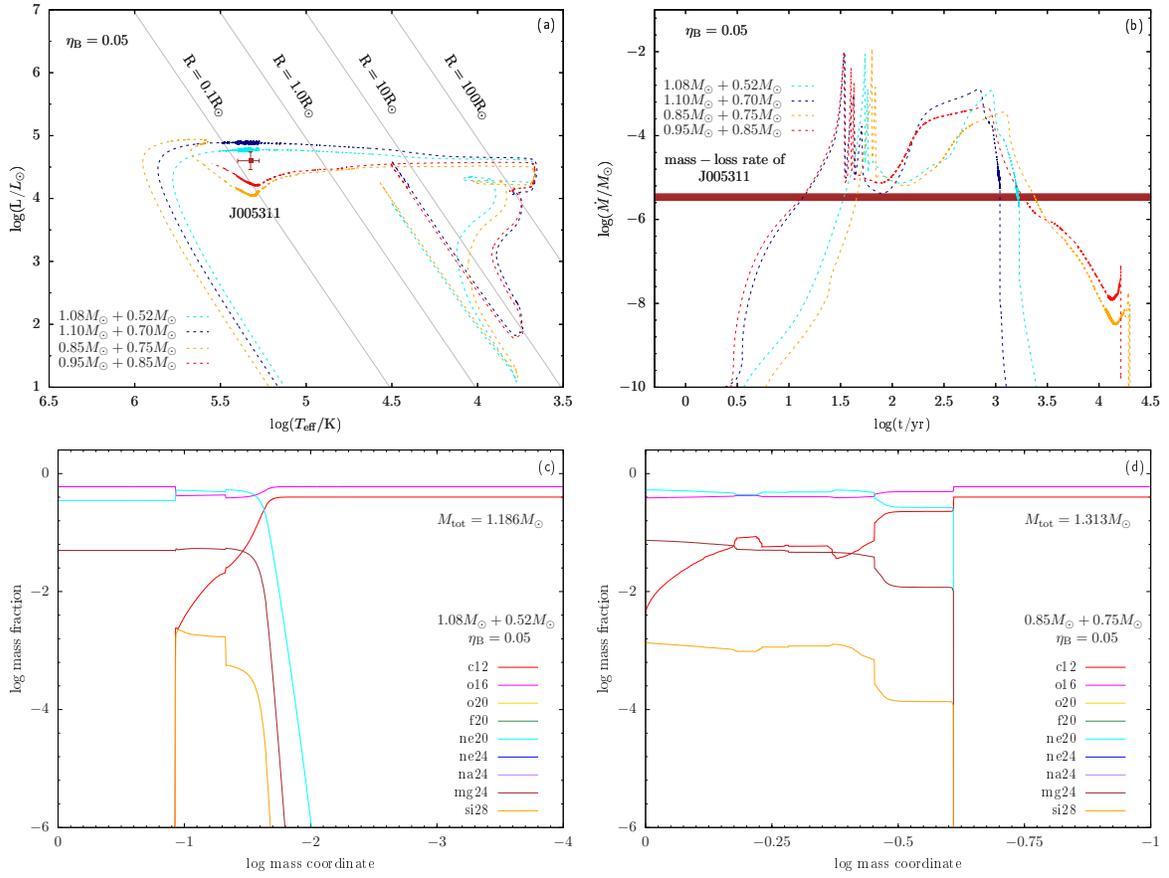,angle=0,width=16.2cm}
 \caption{Evolution of ONe+CO WD and CO+CO WD models when considering rotation (${\Omega}_{\rm c}=0.5\,{\rm rad}/{\rm s}$, ${\Omega}_{\rm e}=0.5{\Omega}_{\rm crit}$) and Blocker's wind mass-loss prescription (${\eta}_{\rm B}=0.05$). Panel (a): HRD. Cyan and navy lines represent evolution of ONe+CO WD merger remnants with total masses equal to $1.60{M}_{\odot}$ and $1.80{M}_{\odot}$, respectively. In comparison, the evolution of double CO WD mergers with same total masses (mass ratio equal to 0.9) are shown as orange and red lines, respectively. Panel (b): evolution of wind mass-loss rate for the corresponding models. Brown point with error bar represents the position of J005311 on HRD, whereas the brown shaded area in panel (b) represents the estimated wind mass-loss rate of J005311. Thick solid portion of each lines in panel (a) and (b) represent the position of each models when they evolve to the same temperature as J005311. Panel (c) and (d): Elemental abundance distributions of ONe+CO WD and CO+CO WD models at the moment when they evolve to the position close to the object of J005311 on HRD, respectively.}
  \end{center}
    \label{fig: 10}
\end{figure*}

\begin{table*}[htbp]
\centering
\begin{tabular}{|c|c|c|c|c|c|c|}
\hline
 & & & & & & \\
{Model} & ${M}_{\rm tot}/{M}_{\odot}$ & ${\rm ln}({s}_{\rm env}/{\rm erg}\,{\rm g}^{-1}\,{\rm K}^{-1})$ & ${T}_{\rm max}/{10}^{8}\,{\rm K}$ & Stopping condition & ${\rm t}/{\rm yr}$ & Expected fate \\
 & & & & & & \\
\hline
ONe+CO & & & & & & \\
\hline
$1.10+0.65$ & $1.75$ & $8.875$ & $10.484$ & ${\rm log}({T}_{\rm c}/{\rm K})>9.34$ & $9687$ & NS \\
\hline
$1.10+0.70$ & $1.80$ & $8.890$ & $10.788$ & ${\rm log}({T}_{\rm c}/{\rm K})>9.34$ & $8173$ & NS \\
$1.15+0.65$ & $1.80$ & $8.890$ & $11.660$ & ${\rm log}({T}_{\rm c}/{\rm K})>9.34$ & $5681$ & NS \\
\hline
$1.10+0.75$ & $1.85$ & $8.905$ & $11.138$ &  & $6202$ & NS \\
$1.15+0.70$ & $1.85$ & $8.905$ & $11.929$ & ${\rm log}({T}_{\rm c}/{\rm K})>9.34$ & $4112$ & NS \\
$1.20+0.65$ & $1.85$ & $8.905$ & $13.114$ &  & $994$ &  NS \\
\hline
$1.10+0.80$ & $1.90$ & $8.920$ & $11.579$ &  & $3125$ & NS \\
$1.15+0.75$ & $1.90$ & $8.920$ & $12.339$ & ${\rm log}({T}_{\rm c}/{\rm K})>9.34$ & $1729$ & NS \\
$1.20+0.70$ & $1.90$ & $8.920$ & $13.398$ &  & $178$ &  ONeFe WD \\
\hline
$1.10+0.85$ & $1.95$ & $8.935$ & $12.154$ &  & $240$ &  ONeFe WD \\
$1.15+0.80$ & $1.95$ & $8.935$ & $12.782$ & ${\rm log}({T}_{\rm c}/{\rm K})>9.34$ & $179$ &  ONeFe WD \\
$1.20+0.75$ & $1.95$ & $8.935$ & $13.824$ &  & $107$ &  ONeFe WD \\
\hline
$1.10+0.90$ & $2.00$ & $8.950$ & $12.649$ &  & $134$ &  ONeFe WD \\
$1.15+0.85$ & $2.00$ & $8.950$ & $13.242$ & ${\rm log}({T}_{\rm c}/{\rm K})>9.34$ & $163$ &  ONeFe WD \\
$1.20+0.80$ & $2.00$ & $8.950$ & $14.271$ &  & $47$ &   ONeFe WD \\
\hline
$1.10+0.95$ & $2.05$ & $8.965$ & $13.330$ &  & $106$ &  ONeFe WD \\
$1.15+0.90$ & $2.05$ & $8.965$ & $13.850$ & ${\rm log}({T}_{\rm c}/{\rm K})>9.34$ & $66$ &   ONeFe WD \\
$1.20+0.85$ & $2.05$ & $8.965$ & $14.771$ &  & $39$ &   ONeFe WD \\
\hline
$1.10+1.00$ & $2.10$ & $8.980$ & $14.053$ &  & $59$ &  ONeFe WD \\
$1.15+0.95$ & $2.10$ & $8.980$ & $14.556$ & ${\rm log}({T}_{\rm c}/{\rm K})>9.34$ & $43$ &  ONeFe WD \\
$1.20+0.90$ & $2.10$ & $8.980$ & $15.387$ &  & $25$ &  ONeFe WD \\
\hline
$1.10+1.05$ & $2.15$ & $8.995$ & $14.959$ &  & $--$ &  ONeFe WD \\
$1.15+1.00$ & $2.15$ & $8.995$ & $15.486$ & ${\rm log}({T}_{\rm c}/{\rm K})>9.34$ & $51$ &  ONeFe WD \\
$1.20+0.95$ & $2.15$ & $8.995$ & $16.234$ &  & $20$ &  ONeFe WD \\
\hline
$1.15+1.05$ & $2.20$ & $9.010$ & $16.441$ & ${\rm log}({T}_{\rm c}/{\rm K})>9.34$ & $232$ & ONeFe WD \\
$1.20+1.00$ & $2.20$ & $9.010$ & $10.720$ & ${\rm log}({T}_{\rm c}/{\rm K})>9.34$ & $14$ &  ONeFe WD \\
\hline
$1.20+1.05$ & $2.25$ & $9.025$ & $18.350$ & ${\rm log}({T}_{\rm c}/{\rm K})>9.34$ & $8.4$ & ONeFe WD \\
\hline
\end{tabular}
\caption{\label{tab:3.2} Models involved in sect.\,3.2. Columns 1-7 represent, (1) Model: masses of ONe WD and CO WD; (2) ${M}_{\rm tot}/{M}_{\odot}$: total mass of the merger remnant; (3) ${\rm ln}({s}_{\rm env}/{\rm erg}\,{\rm g}^{-1}\,{\rm K}^{-1})$: natural logarithm of envelope entropy of remnant; (4) ${T}_{\rm max}/{10}^{8}\,{\rm K}$: maximum temperature of remnant; (5) Stopping condition: stage at which calculation is stopped; (6) ${\rm t}/{\rm yr}$: evolution times from the beginning of our calculations to the stopping conditions (the model of $1.10+1.05$ meets some numerical problems during its evolution); (7) Expected fate: final evolutionary outcome of the remnant. NS represents that the remnant experiences ECSN to forms neutron star, whereas ONeFe WD represents that the remnant experiences ECSN to forms ONeFe WD.} 
\end{table*}

\begin{table*}[htbp]
\centering
\begin{tabular}{|c|c|c|c|c|c|c|}
\hline
 & & & & & & \\
{Model} & ${M}_{\rm tot}/{M}_{\odot}$ & ${\rm ln}({s}_{\rm env}/{\rm erg}\,{\rm g}^{-1}\,{\rm K}^{-1})$ & ${T}_{\rm max}/{10}^{8}\,{\rm K}$ & Stopping condition & ${\rm t}/{\rm yr}$ & Expected fate \\
 & & & & & & \\
\hline
ONe+CO & & & & & & \\
\hline
$1.10+0.80$ & $1.90$ & $8.70$ & $15.133$ & ${\rm log}({T}_{\rm c}/{\rm K})>9.34$ & $80$ & ONeFe WD \\
$1.10+0.80$ & $1.90$ & $9.10$ & $9.773$ & ${\rm log}({T}_{\rm c}/{\rm K})>9.34$ & $10963$ & NS \\
\hline
$1.10+0.85$ & $1.95$ & $8.70$ & $17.034$ & ${\rm log}({T}_{\rm c}/{\rm K})>9.34$ & $45.5$ & ONeFe WD \\
$1.10+0.85$ & $1.95$ & $9.10$ & $10.228$ & ${\rm log}({T}_{\rm c}/{\rm K})>9.34$ & $8788$ & NS \\
\hline
\end{tabular}
\caption{\label{tab:4.1} Models involved in sect.\,4.1. Columns 1-7 represent, (1) Model: masses of ONe WD and CO WD; (2) ${M}_{\rm tot}/{M}_{\odot}$: total mass of the merger remnant; (3) ${\rm ln}({s}_{\rm env}/{\rm erg}\,{\rm g}^{-1}\,{\rm K}^{-1})$: natural logarithm of envelope entropy of remnant; (4) ${T}_{\rm max}/{10}^{8}\,{\rm K}$: maximum temperature of remnant; (5) Stopping condition: stage at which calculation is stopped; (6) ${\rm t}/{\rm yr}$: evolution times from the beginning of our calculations to the stopping conditions; (7) Expected fate: final evolutionary outcome of the remnant.}  
\end{table*}

\begin{table*}[htbp]
\centering
\begin{tabular}{|c|c|c|c|c|c|}
\hline
 & & & & & \\
{Model} & ${M}_{\rm tot}/{M}_{\odot}$ & Mass-loss prescription & Stopping condition & ${\rm t}/{\rm yr}$ & Expected fate \\
 & & & & & \\
\hline
ONe+CO & & & & & \\
\hline
$1.10+0.80$ & $1.90$ & ${\eta}_{\rm R}=0.5$ & ${\rm log}({T}_{\rm c}/{\rm K})>9.34$ & $3121$ & NS \\
$1.10+0.80$ & $1.90$ & ${\eta}_{\rm B}=0.05$ & ${\rm log}({L}/{L}_{\odot})<1.0$ & $29960$ & ONe WD \\
\hline
$1.10+0.85$ & $1.95$ & ${\eta}_{\rm R}=0.5$ & ${\rm log}({T}_{\rm c}/{\rm K})>9.34$ & $242$ & ONeFe WD \\
$1.10+0.85$ & $1.95$ & ${\eta}_{\rm B}=0.05$ & ${\rm log}({T}_{\rm c}/{\rm K})>9.34$ & $241$ & ONeFe WD \\
\hline
\end{tabular}
\caption{\label{tab:4.2} Models involved in sect.\,4.2. Columns 1-6 represent, (1) Model: masses of ONe WD and CO WD; (2) ${M}_{\rm tot}/{M}_{\odot}$: total mass of the merger remnant; (3) Mass-loss prescription: wind mass-loss prescription considered in the corresponding model. ``${\eta}_{\rm R}=0.5$'' represents models adopt Reimers' wind mass-loss prescription with factor of ${\eta}_{\rm R}=0.5$, whereas ``${\eta}_{\rm B}=0.05$'' represents models adopt Blocker's wind mass-loss prescription with factor of ${\eta}_{\rm B}=0.05$; (4) Stopping condition: stage at which calculation is stopped; (5) ${\rm t}/{\rm yr}$: evolution times from the beginning of our calculations to the stopping conditions; (6) Expected fate: final evolutionary outcome of the remnant. NS represents that the remnant experiences ECSN to forms neutron star, whereas ONeFe WD represents that the remnant experiences ECSN to forms ONeFe WD.}  
\end{table*}

\begin{table*}[htbp]
\centering
\begin{tabular}{|c|c|c|c|c|c|}
\hline
 & & & & & \\
{Model} & ${M}_{\rm tot}/{M}_{\odot}$ & ${f}_{\rm He}$ in envelope & Stopping condition & ${\rm t}/{\rm yr}$ & Expected fate \\
 & & & & & \\
\hline
ONe+CO & & & & & \\
 \hline
$1.10+0.50$ & $1.60$ & $0\%$ & ${\rm log}({T}_{\rm c}/{\rm K})>9.34$ & $12507$ & NS \\
\hline
ONe+HeCO & & & & & \\
\hline
$1.10+0.50$ & $1.60$ & $20\%$ & ${\rm log}({T}_{\rm c}/{\rm K})>9.34$ & $27859$ & NS \\
\hline
\end{tabular}
\caption{\label{tab:4.3} Models involved in sect.\,4.3. Columns 1-6 represent, (1) Model: masses of ONe WD and CO WD; (2) ${M}_{\rm tot}/{M}_{\odot}$: total mass of the merger remnant; (3) ${f}_{\rm He}$ in envelope: mass fraction of $^{\rm 4}{\rm He}$ in CO envelope; (4) Stopping condition: stage at which calculation is stopped; (5) ${\rm t}/{\rm yr}$: evolution times from the beginning of our calculations to the stopping conditions; (6) Expected fate: final evolutionary outcome of the remnant. NS represents that the remnant experiences ECSN to forms neutron star.} 
\end{table*}

\begin{table*}[htbp]
\centering
\begin{tabular}{|c|c|c|c|c|c|c|c|c|c|}
\hline
 & & & & & & & & & \\
{Model} & ${M}_{\rm tot}/{M}_{\odot}$ & ${f}_{^{\rm 16}{\rm O}}$ & ${f}_{^{\rm 20}{\rm Ne}}$ & ${f}_{^{\rm 24}{\rm Mg}}$ & ${f}_{^{\rm 23}{\rm Na}}$ & ${f}_{^{\rm 25}{\rm Mg}}$ & Stopping condition & ${\rm t}/{\rm yr}$ & Expected fate \\
 & & & & & & & & & \\
\hline
ONe+CO & & & & & & & & & \\
\hline
$1.10+0.80$ & $1.90$ & $60\%$ & $35\%$ & $5\%$ & $0\%$ & $0\%$ & ${\rm log}({T}_{\rm c}/{\rm K})>9.34$ & $3125$ & NS \\
$1.10+0.80$ & $1.90$ & $54\%$ & $35\%$ & $5\%$ & $5\%$ & $1\%$ & ${\rm log}({T}_{\rm c}/{\rm K})>9.34$ & $1937$ & NS \\
\hline
$1.10+0.85$ & $1.95$ & $60\%$ & $35\%$ & $5\%$ & $0\%$ & $0\%$ & ${\rm log}({T}_{\rm c}/{\rm K})>9.34$ & $240$ & ONeFe WD \\
$1.10+0.85$ & $1.95$ & $54\%$ & $35\%$ & $5\%$ & $5\%$ & $1\%$ & ${\rm log}({T}_{\rm c}/{\rm K})>9.34$ & $197$ & ONeFe WD \\
\hline
\end{tabular}
\caption{\label{tab:4.4} Models involved in sect.\,4.4. Columns 1-10 represent, (1) Model: masses of ONe WD and CO WD; (2) ${M}_{\rm tot}/{M}_{\odot}$: total mass of the merger remnant; (3) ${f}_{^{\rm 16}{\rm O}}$: mass fraction of $^{\rm 16}{\rm O}$ in ONe core; (4) ${f}_{^{\rm 20}{\rm Ne}}$: mass fraction of $^{\rm 20}{\rm Ne}$ in ONe core; (5) ${f}_{^{\rm 24}{\rm Mg}}$: mass fraction of $^{\rm 24}{\rm Mg}$ in ONe core; (6) ${f}_{^{\rm 23}{\rm Na}}$: mass fraction of $^{\rm 23}{\rm Na}$ in ONe core; (7) ${f}_{^{\rm 25}{\rm Mg}}$: mass fraction of $^{\rm 25}{\rm Mg}$ in ONe core; (8) Stopping condition: stage at which calculation is stopped; (9) ${\rm t}/{\rm yr}$: evolution times from the beginning of our calculations to the stopping conditions; (10) Expected fate: final evolutionary outcome of the remnant. NS represents that the remnant experiences ECSN to forms neutron star, whereas ONeFe WD represents that the remnant experiences ECSN to forms ONeFe WD.} 
\end{table*}

\begin{table*}[htbp]
\centering
\begin{tabular}{|c|c|c|c|c|c|c|c|}
\hline
 & & & & & & & \\
{Model} & ${M}_{\rm tot}/{M}_{\odot}$ & ${\Omega}_{\rm c}/{\rm rad}\,{\rm s}^{-1}$ & ${\Omega}_{\rm e}/{\Omega}_{\rm crit}$ & AM transport & Stopping condition & ${\rm t}/{\rm yr}$ & Expected fate \\
 & & & & & & & \\
\hline
ONe+CO & & & & & & & \\
 \hline
$1.10+0.80$ & $1.90$ & $0.50$ & $0.50$ & Set 1 & ${\rm log}({T}_{\rm c}/{\rm K})>9.34$ & $3020$ & NS \\
$1.10+0.80$ & $1.90$ & $0.50$ & $0.50$ & Set 2 & ${\rm log}({T}_{\rm c}/{\rm K})>9.34$ & $3050$ & NS \\
\hline
$1.10+0.80$ & $1.90$ & $0.25$ & $0.25$ & Set 1 & ${\rm log}({T}_{\rm c}/{\rm K})>9.34$ & $3127$ & NS \\
$1.10+0.80$ & $1.90$ & $0.25$ & $0.25$ & Set 2 & ${\rm log}({T}_{\rm c}/{\rm K})>9.34$ & $3135$ & NS \\
\hline
$1.10+0.85$ & $1.95$ & $0.50$ & $0.50$ & Set 1 & ${\rm log}({T}_{\rm c}/{\rm K})>9.34$ & $243$ & ONeFe WD \\
$1.10+0.85$ & $1.95$ & $0.50$ & $0.50$ & Set 2 & ${\rm log}({T}_{\rm c}/{\rm K})>9.34$ & $214$ & ONeFe WD \\
\hline
$1.10+0.85$ & $1.95$ & $0.25$ & $0.25$ & Set 1 & ${\rm log}({T}_{\rm c}/{\rm K})>9.34$ & $243$ & ONeFe WD \\
$1.10+0.85$ & $1.95$ & $0.25$ & $0.25$ & Set 2 & ${\rm log}({T}_{\rm c}/{\rm K})>9.34$ & $245$ & ONeFe WD \\
\hline
\end{tabular}
\caption{\label{tab:4.5} Models involved in sect.\,4.5. Columns 1-8 represent, (1) Model: masses of ONe WD and CO WD; (2) ${M}_{\rm tot}/{M}_{\odot}$: total mass of the merger remnant; (3) ${\Omega}_{\rm c}/{\rm rad}\,{\rm s}^{-1}$: rotation velocity of core in ${\rm rad}/{\rm s}$; (4) ${\Omega}_{\rm e}/{\Omega}_{\rm crit}$: rotation velocity of each mass zone in envelope divided by Keplerian velocity of the corresponding mass zone; (5) AM transport: angular momentum transport prescription adopted in the corresponding models; Set 1 represents adopting the prescription described in Fuller et al. (2019), whereas Set 2 represents which described in Spruit (2002); (6) Stopping condition: stage at which calculation is stopped; (7) ${\rm t}/{\rm yr}$: evolution times from the beginning of our calculations to the stopping conditions; (8) Expected fate: final evolutionary outcome of the remnant. NS represents that the remnant experiences ECSN to forms neutron star, whereas ONeFe WD represents that the remnant experiences ECSN to forms ONeFe WD.} 
\end{table*}

\begin{table*}[htbp]
\centering
\begin{tabular}{|c|c|c|c|c|c|c|c|}
\hline
 & & & & & & & \\
{Model} & ${M}_{\rm tot}/{M}_{\odot}$ & ${\Omega}_{\rm c}/{\rm rad}\,{\rm s}^{-1}$ & ${\Omega}_{\rm e}/{\Omega}_{\rm crit}$ & Mass-loss prescription & Stopping condition & ${\rm t}/{\rm yr}$ & Expected fate \\
 & & & & & & & \\
\hline
ONe+CO & & & & & & & \\
\hline
$1.08+0.52$ & $1.60$ & $0.50$ & $0.50$ & ${\eta}_{\rm B}=0.05$ & ${\rm log}({L}/{L}_{\odot})<1.0$ & $86306$ & ONe WD \\
$1.10+0.70$ & $1.80$ & $0.50$ & $0.50$ & ${\eta}_{\rm B}=0.05$ & ${\rm log}({L}/{L}_{\odot})<1.0$ & $56030$ & ONe WD \\
\hline
CO+CO & & & & & & & \\
\hline
$0.85+0.75$ & $1.60$ & $0.50$ & $0.50$ & ${\eta}_{\rm B}=0.05$ & ${\rm log}({L}/{L}_{\odot})<1.0$ & $52378$ & ONe WD \\
$0.95+0.85$ & $1.80$ & $0.50$ & $0.50$ & ${\eta}_{\rm B}=0.05$ & Ne ignition & $16276$ & Fe-CCSN \\
\hline
\end{tabular}
\caption{\label{tab:5} Models involved in sect.\,5. Columns 1-8 represent, (1) Model: masses of ONe WD and CO WD; (2) ${M}_{\rm tot}/{M}_{\odot}$: total mass of the merger remnant; (3) ${\Omega}_{\rm c}/{\rm rad}\,{\rm s}^{-1}$: rotation velocity of core in ${\rm rad}/{\rm s}$; (4) ${\Omega}_{\rm e}/{\Omega}_{\rm crit}$: rotation velocity of each mass zone in envelope divided by Keplerian velocity of the corresponding mass zone; (5) Mass-loss prescription: wind mass-loss prescription considered in the corresponding model. ``${\eta}_{\rm B}=0.05$'' represents models adopt Blocker's wind mass-loss prescription with factor of ${\eta}_{\rm B}=0.05$; (6) Stopping condition: stage at which calculation is stopped; (7) ${\rm t}/{\rm yr}$: evolution times from the beginning of our calculations to the stopping conditions; (8) Expected fate: final evolutionary outcome of the remnant. ONe WD represents that the remnant end its evolution as ONe WD, whereas Fe-CCSN represents that the remnant experiences Fe-CCSN to forms NS.}
\end{table*}

\section{Summary}

In this work, we investigated the evolution of ONe WD+CO WD merger remnants for the first time. We found that the evolutionary outcomes of such merger remnants are related to their initial masses, i.e. less massive remnants (${M}\leq1.90{M}_{\odot}$) would experience shell carbon burning phase to increase the core mass before off-center neon burning, whereas more massive remnants (${M}\geq1.95{M}_{\odot}$) would trigger off-center neon burning soon after the merger. By considering wind mass-loss prescriptions, the less massive remnants show more sensitive to the wind mass-loss rate, resulting in that their final fates may be ONe WDs or NSs. By contrast, the more massive remnants may always experience supernova explosions to form ONeFe WDs. Using the mass-loss prescription by \cite{1995A&A...297..727B}, our models can explain the observational properties of J005311, including temperature, luminosity, wind mass-loss rate, the formation of a nebula, the evolutionary time, and the abundances of the ejecta. From this, We can derive, first that our study may be able to provide some constraints on the wind mass-loss rates of H-/He-deficient giant-like objects similar to J005311. For example, due to the relevance of dust in both environments, their wind mass-loss process could present analogies with that occurring in AGB stars. Finally, we strongly suggest that J005311 originates from the merger of an ONe WD with a CO WD, and the final fate of this object would be to become a massive ONe WD rather than supernova explosion.

\begin{acknowledgments}
We thank Matthias Kruckow, Xiangcun Meng, Zhengwei Liu, Jiangdan Li, Hailiang Chen and Dongdong Liu for helpful discussion. This study is supported by the National Key R\&D Program of China (No. 2021YFA1600404), the National Natural Science Foundation of China (NSFC grants 12003013, 12225304, 12288102 and 12033003), the Western Light Project of CAS (No. XBZG-ZDSYS-202117), the science research grants from the China Manned Space Project (No CMS-CSST-2021-A12), the Yunnan Fundamental Research Projects (No. 202001AS070029), the Scholar Program of Beijing Academy of Science and Technology (DZ:BS202002), and the Tencent Xplorer Prize.
\end{acknowledgments}

\appendix

The source files for our MESA models are publicly available on Zenodo at https://doi.org/10.5281/zenodo.7456455.


\end{document}